\documentclass[referee]{raa}            
\usepackage{graphicx}	
\usepackage{amsmath}	
\usepackage{siunitx}
\usepackage{amssymb}
\usepackage{gensymb}
\usepackage{color}
\usepackage{natbib}
\usepackage{enumitem}
\usepackage{mathtools}
\usepackage{lipsum}
\usepackage{threeparttable}
\usepackage{textcomp,booktabs}
\usepackage{multirow}
\usepackage{verbatim}
\usepackage{rotating}
\usepackage{multicol}        
\usepackage{bm}		
\usepackage{pdflscape}	
\usepackage{txfonts}
\usepackage{subcaption}
\usepackage[T1]{fontenc}
\usepackage{ae,aecompl}
\bibpunct{(}{)}{;}{a}{}{,}
\usepackage[pagebackref=true]{hyperref}

\begin{document}

   \title{The spin measurement of MAXI J0637-430: a black hole candidate with high disk density}

 \volnopage{ {\bf 20XX} Vol.\ {\bf X} No. {\bf XX}, 000--000}
   \setcounter{page}{1}

   \author{Nan Jia\inst{1,2}, Ye Feng\inst{1,2}, Yu-jia Song\inst{1,2},  Jun Yang\inst{1,2}, Jieun Yuh\inst{3} , Pei-jun Huang\inst{4} and Li-jun Gou\inst{1,2}\thanks{E-mail: lgou@nao.cas.cn} 
   }

   \institute{ Key Laboratory for Computational Astrophysics, National Astronomical Observatories, Chinese Academy of Sciences,  Datun Road 20A, Beijing 100101, China; {\it nanjia@nao.cas.cn}\\
        \and
             School of Astronomy and Space Sciences, University of Chinese Academy of Sciences,  Yuquan Road 19A, Beijing 100049, China\\
          \and
             Shanghai American School Puxi Campus, 258 Jinfeng Road, Shanghai 201107, China\\
           \and
             Shenzhen Yaohua Experimental School, No.99 JingTian South 5th Street, Futian District, Shenzhen 518034, China\\
\vs \no
   {\small Received 20XX Month Day; accepted 20XX Month Day}
}

\abstract{The Galactic black hole candidate MAXI J0637-430 was first discovered by \textit{MAXI/GSC} on 2019 November 02. We study the spectral properties of MAXI J0637-430 by using the archived \textit{NuSTAR} data and \textit{Swift}/XRT data.  After fitting the eight spectra by using a disk component and a powerlaw component model with absorption, we select the spectra with relatively strong reflection components for detailed X-ray reflection spectroscopy. Using the most state-of-art reflection model $\tt{relxillCp}$, the spectral fitting measures a black hole spin $\textit{a}_{\rm{*}} >  0.72$ and the inclination angle of the accretion disk $i$ = $46.1_{-5.3}^{+4.0}$ degrees, at 90 per cent confidence level. In addition, the fitting results show an extreme supersolar iron abundance.
Combined with the fitting results of the reflection model $\tt{reflionx\_hd}$, we consider that this unphysical iron abundance may be caused by a very high density accretion disk ( $n_{\rm{e}} > 2.34 \times 10^{21}$ $\rm{cm}^{-3}$ ) or a strong Fe K$\alpha$ emission line.
The soft excess is found in the soft state spectral fitting results, which may be an extra free-free heating effect caused by high density of the accretion disk.
Finally, we discuss the robustness of black hole spin obtained by X-ray reflection spectroscopy.
The result of relatively high spin is self-consistent with broadened Fe K$\alpha$ line.
Iron abundance and disk density have no effect on the spin results.
}

   \authorrunning{Nan Jia et al. }            
   \titlerunning{Spin measurement of MAXI J0637-430}  
   \maketitle

\section{Introduction}           
\label{sec1}

An X-ray binary consists of a compact object and a donor star. According to the donor star mass, the X-ray binary can be divided into High mass X-ray binary (HMXB) and Low mass X-ray binary (LMXB). 
In LMXBs, the companion star fills the Roche lobe and forms an accretion disk around the compact object. 
While in HMXBs, the accreted matters from the companion stars via wind.
The compact object in the X-ray binary is commonly found to be a neutron star or a stellar-mass black hole. 
So far, at least 20 black hole X-ray binaries (BHXRBs) have been discovered \citep{reynolds2021observational}. 
Most of the BHXRBs are transients, and a few are persistents such as Cyg X-1 and LMC X-1.

BHXRBs are ideal objects for testing general relativity and studying the physical properties of black holes with surrounding structures, like the accretion disk and corona. 
For a real astrophysical environment, a black hole can be characterized by the black hole mass ($M_{\rm{BH}}$) and the black hole spin ($a_*$). 
Usually the black hole mass can be measured from dynamical studies, including mass function and radial velocity curve \citep{orosz2002dynamical,orosz200715}.
As for the black hole spin measurement, it could be more complicated.
Since the BHXRBs show various X-ray spectral features during the whole outburst, which means that the geometric structure of the accretion disk and the physical properties of the corona have been changed.
The black hole X-ray binaries will experience an evolution from hard state (HS) to soft state (SS), with a short time intermediate state between them \citep{remillard2006x}.
It is generally believed that the inner disk radius extends to the innermost stable circular orbit (ISCO) in the soft state.
According to \citet{bardeen1972rotating}, there is a degenerate relationship between the ISCO radius and black hole spin.
So once we have obtained the inner disk radius, we can estimate the black hole spin via X-ray spectroscopy.
At present, two methods that are widely used to measure the black hole spin are the continuum-fitting method, which models the profile of the thermal emission from the accretion disk \citep{zhang1997black}; and the X-ray reflection spectroscopy method, which models the relativistic broadened Fe K$\alpha$ emission line and Compton hump \citep{fabian1989x}. 
Using the continuum-fitting method to measure the black hole spin requires a prior information, that is the black hole mass, inclination, and distance. 
In most cases, these three dynamical parameters of the black hole are unknown, we need to utilize the other method, the X-ray reflection spectroscopy method.
The thermal radiation emitted from the accretion disk in the vicinity of the black hole undergoes Comptonization in the corona, and the produced power-law radiation will be irradiated back to the accretion disk to produce reflection emission.
The significant reflection features are Fe K$\alpha$ emission line and Compton hump.
At the inner region of the accretion disk in the vicinity of the black hole, the Fe K$\alpha$ emission line is distorted and broadened due to the Doppler effect, beaming effect and gravitational redshift.
Using the reflection model to fit the Fe K$\alpha$ profile and the Compton hump, the black hole spin can be obtained.
So far, several sources have used both two methods to measure the black hole spin, which include Cyg X-1 \citep{gou2014confirmation,zhao2020confirming,zhao2021re,tomsick2013reflection}, XTE J1550-564 \citep{steiner2011spin}, LMC X-1 \citep{gou2009determination,steiner2012broad}, 4U 1543-47\citep{shafee2006estimating,dong2020spin}, GRO J1655-40 \citep{shafee2006estimating,reis2009determining}, GRS 1915+105 \citep{reid2014parallax,miller2013nustar} and GX339-4 \citep{kolehmainen2010limits,garcia2015x}.
Because of the absence of the dynamical parameters or the Fe K$\alpha$ emission line, other sources could only use one of the methods to measure the black hole spin.
BHXRBs such as A0620-00 \citep{gou2010spin}, MAXI J1820+070 \citep{zhao2021estimating}, MAXI J1659-152 \citep{feng2022estimating} and MAXI J1305-704 \citep{feng2022using} have successfully used the continuum-fitting method to obtain the black hole spin.
The X-ray reflection spectroscopy method is also widely used in the study of measuring the black hole spin, like MAXI J1535-571 \citep{xu2018reflection,dong2022analysis}, XTE J1752-223 \citep{garcia2018reflection}, MAXI J1836-194 \citep{dong2020detailed}, AT2019wey \citep{feng2022spectral}, MAXI J1348-630 \citep{jia2022detailed} and MAXI J1803+298 \citep{feng2022spin}.

MAXI J0637-430 is a new transient source \citep{kennea2019maxi}, that was discovered by the Monitor of All-sky X-ray Image Gas Slit Camera (\textit{MAXI/GSC}; \citealt{matsuoka2009maxi}) on November 2nd, 2019 (MJD 58789).
And then the X-ray outburst was detected by several X-ray satellites, such as the Neil Gehrels Swift Observatory X-ray Telescope (\textit{Swift/XRT}; \citealt{burrows2005swift}), the Neutron star Interior Composition Explorer (\textit{NICER}; \citealt{gendreau2012neutron}),  \textit{Insight}-HXMT \citep{apjlac7711bib58}, the Nuclear Spectroscopic Telescope Array (\textit{NuSTAR}; \citealt{harrison2013nuclear}), and the \textit{AstroSAT} \citep{singh2014astrosat}.
The whole outburst lasted about six months.
Unlike the spectral evolution shown by a typical transient, MAXI J0637-430 lacks the characteristics of the hard state at the beginning of the outburst or the duration of hard state is very short \citep{tetarenko2021using}.

Since the outburst in 2019, MAXI J0637-430 has been studied several times.
In the observation of the optical band, the optical counterpart was first observed by the Southern Astrophysical Research (\textit{SOAR}) telescope on November 3rd, 2019 and then observed by \textit{Gemini} in December, 2019 \citep{tetarenko2021using}.
A correlation between the X-ray irradiation heating the accretion disk and the evolution of the He {\sc ii} 4686 \AA\ emission line profiles detected in the optical spectra have been found in \citet{tetarenko2021using}. 
Several research were carried out in the X-ray band.
\citet{jana2021nicer} presented detailed studies of MAXI J0637-430 using by \textit{NICER} and \textit{Swift}.
In the timing analysis, they found no evidence of quasi-periodic oscillations (QPO) in the power density spectrum (PDS) of the source.
Under the assumption of the source distance of $d$ < 10 kpc, they estimated the mass of black hole to be in the range of 5 - 12 $M_\odot$.
This conclusion was also verified by the work of \citet{baby2021revealing}.
The power spectrum density generated in the 0.01-100 Hz present no QPOs by using \textit{AstroSAT}.
\citet{lazar2021spectral} used \textit{NuSTAR} and \textit{Swift} data to analyze the spectra and timing properties of MAXI J0637-430.
They found that a single multicolour disk component could not be well fitted the soft state spectra, and the fitting results showed that there were at least two components.
They suggested that the additional soft excess is the emission from the plunging region or a reflection component from the blackbody returning radiation with a thermal Comptonization component.
Different from typical X-ray binaries, MAXI J0637-430 has undergone a strange evolution, with a rapid transition at the beginning of the outburst, a lower luminosity and a shorter decay timescale \citep{ma2022peculiar}.
By fitting the soft state spectral, they find that it has deviations from the standard $L_{\rm{disk}}$ $\propto$ $T^4_{\rm{in}}$ relationship, and there may exist additional thermal component.
They propose some accretion disk geometry to explain the scenario, like a hotter blackbody component plus a colder disk component or an ionized outflows plus a disk component.

Although MAXI J0637-430 has obtained a mass estimation, we also need to know the black hole spin value if we could fully characterize a black hole.
For this reason, we utilize X-ray reflection spectroscopy to analyze the archived data of \textit{NuSTAR}.
The state-of-art reflection physical model is used to fit the spectrum to obtain the spin and other parameters of the black hole.

This paper is organized as follows. In Section \ref{sec2}, we describe the observations and data reduction of MAXI J0637-430. In Section \ref{sec3}, we present the spectral analysis results. In Section \ref{sec4}, we discuss the the possible reason of the extreme high supersolar abundance, soft excess found in the soft state and robustness of the high spin value. In Section \ref{sec5}, we summarize the results and present our conclusion.

\section{observations and data reduction}
\label{sec2}
We obtained the daily averaged light curve from \textit{MAXI/GSC}\footnote{http://maxi.riken. jp} \citep{matsuoka2009maxi}.
The hardness ratio plot of MAXI J0637-430 is also shown in the same figure.
At the beginning of the outburst, MAXI J0637-430 seems to be missing the hard state or the time scale of the hard state is very short, and it entered the intermediate state (IMS) when it was found.
The X-ray flux reached the maximum at MJD 58793, and then entered the relatively slow decay phase. 
The source remained the soft state from MJD 58800 to MJD 58858 with a low hardness ratio.
From MJD 58858 to MJD 58880, the source X-ray intensity has a steep declination and a significant increase on hardness ratio, which means the source entered in the intermediate state.
It is worth noting that the properties of the two intermediate states are different when the source is in the first intermediate state and when the source returns to the intermediate state for the second time.
From the research of \citet{jana2021nicer}, we can see that the normalization of the accretion disk component shows a great difference, which is also verified in our fitting results in the Section \ref{sec3}.
After that, the source evolved to low hard state and remained low X-ray luminosity till the end of observations. 
Our classification of spectral states refers to \citet{jana2021nicer}.

\subsection{\it{NuSTAR}}
\label{Nus}
The observations of \textit{NuSTAR} started from MJD 58792 and ended in MJD 58801. 
\textit{NuSTAR} made eight observations during the whole outburst. 
We use the v2.0.0 of NuSTARDAS pipeline with version 202103152 of calibration database\footnote{https://heasarc.gsfc.nasa.gov/docs/heasarc/caldb/caldb\_supported\_missions.html} (CALDB) to process the \textit{NuSTAR} archived data\footnote{https://heasarc.gsfc.nasa.gov/FTP/nustar/data/}.
The NuSTAR source spectra are extracted following the standard procedure\footnote{https://heasarc.gsfc.nasa.gov/docs/nustar/analysis/nustar\_swguide.pdf} provided by \textit{NuSTAR} guide.
We then choose a circle (with $\rm{r} = 120^{\prime \prime}$) centered on the source to extract the source spectra.
The background spectra are extracted by using the same circle size ($\rm{r} = 120^{\prime \prime}$) from a source free region.
Using command \texttt{GRPPHA} in HEASOFT v6.28, the \textit{NuSTAR} data are grouped to have at least 25 photons per energy bin.
The state of the source, the exposure time, and the count rates with the different instruments are listed in Table \ref{obsn}.
\begin{figure}
    \centering
    \includegraphics[width=12cm]{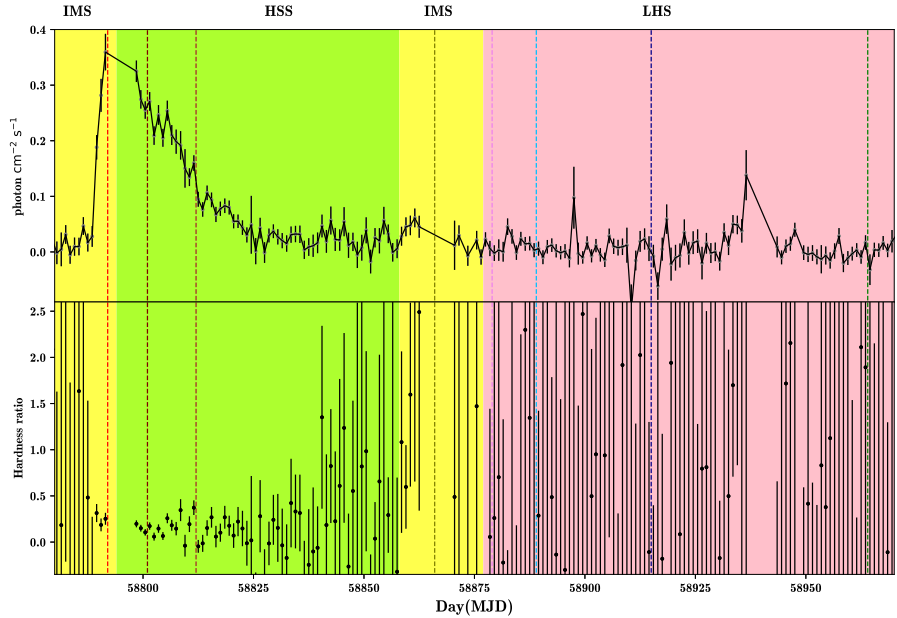}
    \caption{Upper panel: \textit{MAXI/GSC} light curves of MAXI J0637-430 in 2.0 – 20.0 keV. The colourful dashed vertical lines represent the observation of \textit{NuSTAR}. Different spectral states are marked in the color background. Lower panel: time evolution of the hardness ratio (4 – 20 keV/2 – 4 keV). }
    \label{lchid}
\end{figure}

\subsection{\it{Swift}/\rm{XRT}}
When studying the spectrum of X-ray binaries, the disk composition sometimes requires data less than 3 keV to limit the disk temperature, so we process the \textit{Swift}/XRT data to obtain the precise parameters of the thermal component.
The observations of \textit{Swift}/XRT covered whole outburst beginning from MJD 58790.
The \textit{Swift}/XRT spectra over 0.5-10 keV could jointly fit with the \textit{NuSTAR} spectra over 3-79 keV.
However, we notice that the first observation of \textit{NuSTAR}  has no corresponding \textit{Swift}/XRT observation, and the X-ray flux of the  \textit{Swift}/XRT observation (obsID:00088999002) corresponding to the last  \textit{NuSTAR} observation (obsID:80502324016) is too low. 
Here we carry out joint fitting for the second to seventh observations.
The spectra are generated from standard online \textit{Swift}/XRT data products generator provided by UK \textit{Swift} Science Data Centre\footnote{https://www.swift.ac.uk/user\_objects}\citep{evans2009methods}.
All the spectra are grouped to have at least 1 count per bin\citep{kaastra2016optimal}.
Detailed \textit{Swift}/XRT observations are shown in Table \ref{obssw}.

We use \texttt{XSPEC} v12.11.5\footnote{https://heasarc.gsfc.nasa.gov/xanadu/xspec}  to ignore bad channels and then fit all the spectra.
If not specifically mentioned, all uncertainties quoted in this paper are given at 90 per cent confidence level.

\begin{table}
    \caption{\textit{NuSTAR} observation log of MAXI J0637-430}
    \label{obsn}
    \begin{center}
    \setlength{\tabcolsep}{1.2mm}
    \begin{tabular}{cccccc}
			\toprule
                         \toprule
                         ObsID &  MJD & $\rm{State}^{\textit{a}}$ & Instrument & Exposure & $\rm{Count}$ $\rm{rates}^{\textit{b}}$  \\
                         & & & & (ks) & (cts $\rm s^{-1}$)\\
                          \midrule
                        80502324002 & 58792 & IMS & FPMA$\backslash$FPMB& 36.8$\backslash$36.8& 16.89$\backslash$15.20 \\

                   80502324004      &  58801 & SS&  FPMA$\backslash$FPMB  & 67.7$\backslash$67.6 & 11.89$\backslash$10.59\\

                    80502324006      &  58812 &SS &  FPMA$\backslash$FPMB  & 48.6$\backslash$48.4 & 4.67$\backslash$4.12\\

                    80502324008      &  58866 &IMS &  FPMA$\backslash$FPMB  & 46.6$\backslash$46.3 & 2.66$\backslash$2.44\\

                    80502324010      &  58879 &HS &  FPMA$\backslash$FPMB  & 110.8$\backslash$110.0 & 0.77$\backslash$0.73\\

                      80502324012      &  58889&HS &  FPMA$\backslash$FPMB  & 50.2$\backslash$49.9 & 0.39$\backslash$0.37\\

                   80502324014      &  58915 & HS&  FPMA$\backslash$FPMB  & 65.4$\backslash$64.9 & 0.14$\backslash$0.13\\

                   80502324016      &  58964 & HS&  FPMA$\backslash$FPMB  & 47.5$\backslash$47.1 & 0.02$\backslash$0.03\\

                   \bottomrule
    \end{tabular}
    		\begin{tablenotes}
			\item \textbf{Notes.} $^a$The classification of spectral states is referred to \citet{jana2021nicer}.
			
			\qquad\quad$^b$ Count rates are measured in 3.0-79.0 keV for FPMA and FPMB respectively.
		\end{tablenotes}
    \end{center}
\end{table}

\begin{table}
    \caption{\textit{Swift}/XRT observation log of MAXI J0637-430}
    \label{obssw}
    \begin{center}
    \setlength{\tabcolsep}{1.2mm}
    \begin{tabular}{cccccc}
			\toprule
                         \toprule
                         ObsID &  MJD & $\rm{State}^{\textit{a}}$ & Instrument & Exposure & $\rm{Count}$ $\rm{rates}^{\textit{b}}$  \\
                         & & & & (s) & (cts $\rm s^{-1}$)\\
                          \midrule
                        00012172008 & 58801 & SS & XRT& 2515 & 86.70 \\
                      
                      00012172018      &  58812&SS &  XRT  & 1667 & 81.06\\
                      
                        00012172066      &  58866 & IMS& XRT  & 667 & 9.15\\

                    00012172077      &  58879 &HS &  XRT & 1686 & 1.02\\

                    00012172085      &  58889 &HS &  XRT  & $1860^{\textit{c}}$ & $0.36^{\textit{d}}$\\

                    00012172093      &  58915 &HS & XRT  & $944^{\textit{c}}$ &$0.06^{\textit{d}}$\\

                   \bottomrule
    \end{tabular}
    		\begin{tablenotes}
			\item \textbf{Notes.} $^a$The classification of spectral states is referred to \citet{jana2021nicer}.
			
			\qquad\quad$^b$ Count rates of 0.5-10 keV from \textit{Swift}/XRT
			
			\qquad\quad$^c$ Exposure time of the photon counting(PC) mode for the last two observations.
			
			\qquad\quad$^d$ Count rates of the photon counting(PC) mode for the last two observations.
		\end{tablenotes}
    \end{center}
\end{table}

\section{spectral analysis and results}
\label{sec3}
\begin{figure}
    \centering
    \includegraphics[width=10cm,angle=270]{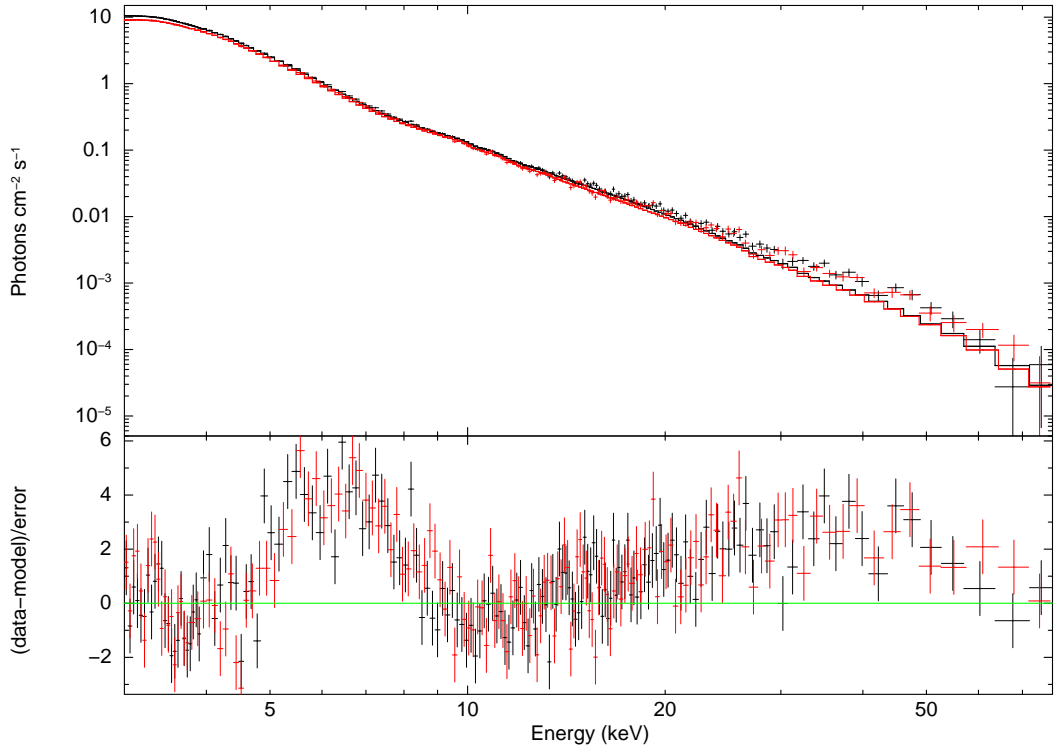}
    \caption{The fitting residuals from model $\tt{constant*tbabs*(diskbb+powerlaw)}$ for 80502324002. FPMA and FPMB data are plotted in black and red, respectively.}
    \label{figmo}
\end{figure}

\begin{figure}
    \centering
    \includegraphics[width=10cm,angle=270]{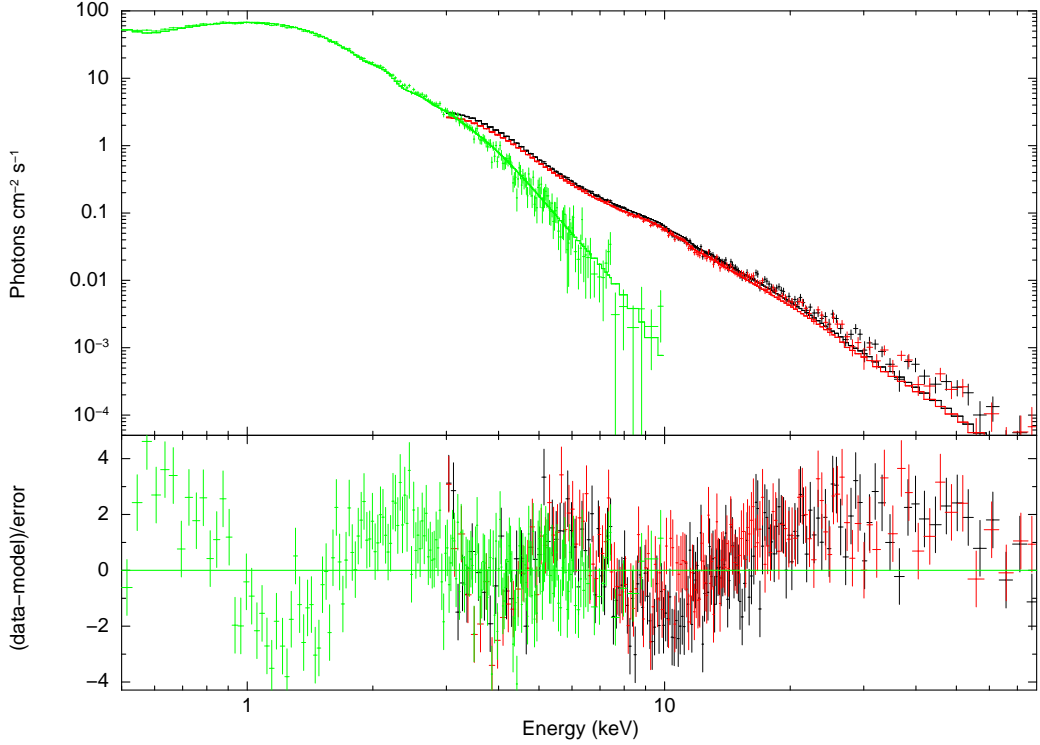}
    \caption{The fitting residuals from model $\tt{constant*tbabs*(diskbb+powerlaw)}$ for 80502324006. XRT, FPMA and FPMB data are plotted in green, black and red, respectively.}
    \label{so}
\end{figure}

In this section, we conduct a detailed spectral analysis of \textit{NuSTAR} and \textit{Swift}/XRT data.
The spectra with strong reflection characteristics are selected to measure the black hole spin.
At the beginning of the fitting process, we use a simple absorbed power-law model \verb'constant*tbabs*powerlaw' to fit the spectra.
\verb'Constant' is used to reconcile the calibration difference among the XRT, FPMA and FPMB. 
When we use the joint fitting of the spectra of NuSTAR and Swift, we fix the constant of XRT and make the constant of FPMA and FPMB change freely.
\verb'tbabs' is the interstellar medium (ISM) absorption model.
We set the cross-sections in \citet{verner1996atomic} and abundances in \citet{wilms2000absorption}.
For the spectra of the intermediate state and the soft state, we add \verb'diskbb' model to fit the accretion disk component.
We fix hydrogen column density (\textit{$N_{\rm{H}}$}) at $4.39\times10^{20}{\rm{cm}^{-2}}$ corresponding to E(B-V) = 0.064 \citep{tetarenko2021using}.
By ignoring 6-7 keV and 15-40 keV to fit the spectra, we find that the spectrum (obsid:80502324002) in the intermediate state has obvious reflection characteristics.
A broadened Fe K$\alpha$ emission line between 6-7 keV and a Compton hump component between 20-50 keV are clearly shown in Fig. \ref{figmo}.
This is the reason that we choose this spectrum to measure the black hole spin.
All the fitting results are listed in Table \ref{po}.
From these fitting results, the spectra of the hard state could well be constrained by model \verb'constant*tbabs*powerlaw' with a good fitting statistics.
However, when using model \verb'constant*tbabs*(diskbb+powerlaw)' to jointly fit the \textit{Swift}/XRT and \textit{NuSTAR} spectra of soft states, it is found that there exist additional residuals in the soft energy bands. 
As shown in Fig. \ref{so}, there is an obvious residual near 1 keV in the spectrum of the soft state.
We will discuss this spectral feature in Section \ref{soex}.

\begin{figure}
    \centering
    \includegraphics[width=10cm,angle=270]{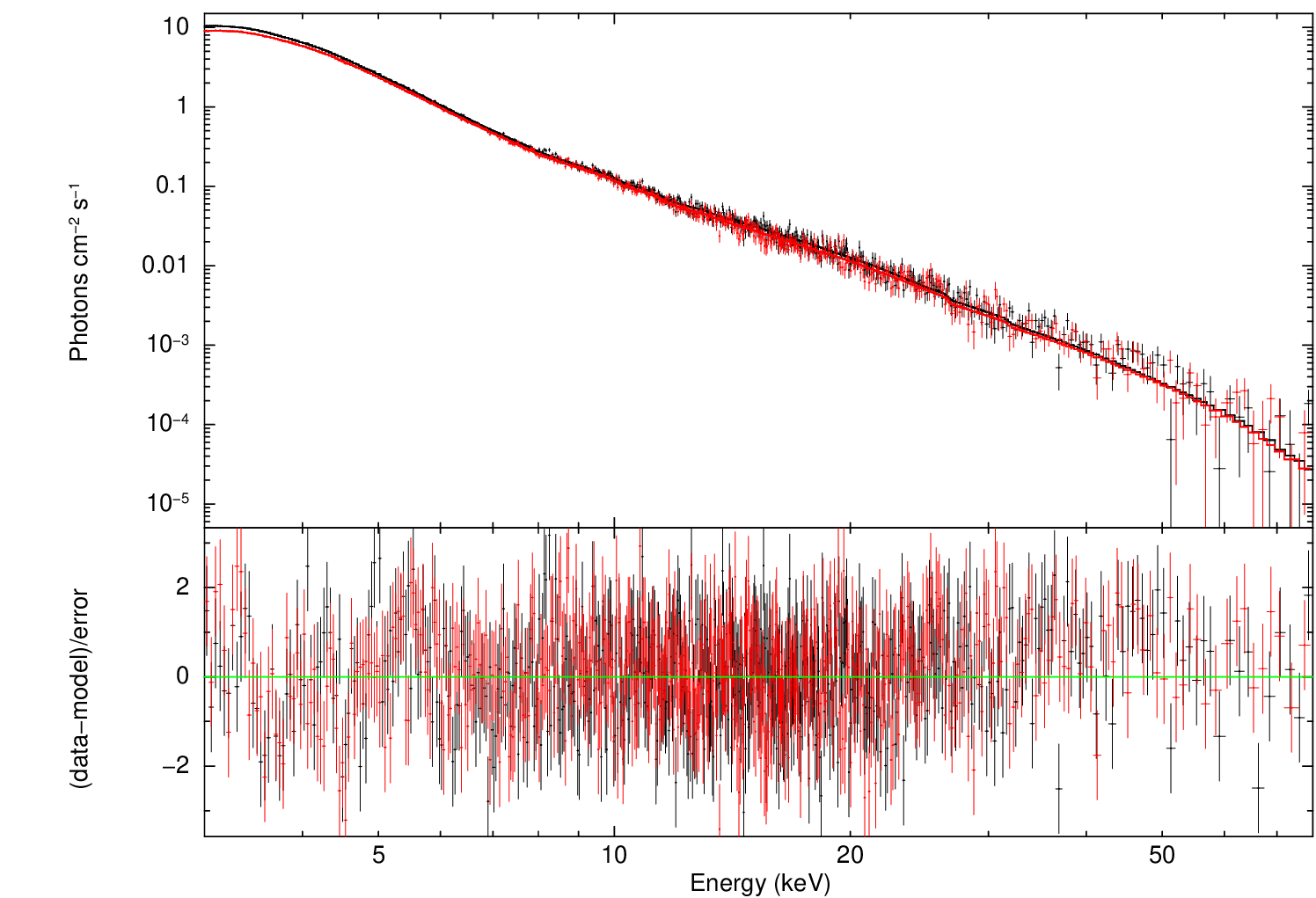}
    \caption{The fitting residuals from model $\tt{constant*tbabs*(diskbb+relxillCp)}$ for 80502324002. FPMA and FPMB data are plotted in black and red, respectively.}
    \label{figre}
\end{figure}

Then, we use a preliminary phenomenological model \verb'constant*tbabs*(diskbb+gaussian+powerlaw)' to fit the spectrum of the first observation.
The model \verb'gaussian' represents the iron emission line and the central energy is set at 6.4 keV.
After adding the model \verb'gaussian', the fitting result has been greatly improved with ${\chi_{\nu}^{2}}$ = 1.20.
In order to measure the spin of MAXI J0637-430, we use the most state-of-art reflection model \verb'relxillCp' v2.2\footnote{http://www.sternwarte.uni-erlangen.de/~dauser/research/relxill} \citep{dauser2014role,garcia2014improved} to fit the spectrum of the intermediate state.
The model \verb'relxillCp' is widely utilized in the research of black hole X-ray binary systems \citep{xu2018reflection,wang2018evolution,sharma2019study}.
The model combines the normal reflection model \verb'xillver' \citep{garcia2010x,garcia2011x,garcia2013x} and the relativistic model \verb'relline' \citep{dauser2010broad,dauser2013irradiation}.
And the incident spectrum in \verb'relxillCp' is the \verb'nthcomp' Comptonization continuum.
The configuration of our physical model is \verb'constant*tbabs*(diskbb+relxillCp)'.
Before measuring the black hole spin, we need to examine the location of the inner radius of the disk ($R_{\rm{in}}$).
If the inner radius of the disk does not extend to the ISCO, the truncation of the accretion disk may occur, thus affecting the measurement of the black hole spin. 
We set the black hole spin ($a_*$) at the maximum value of 0.998 and make the $R_{\rm{in}}$ fitted freely.
For the parameters of the emissivity index, we assume a canonical case ($q_{\rm out}$ = $q_{\rm in}$ = $3$) \citep{fabian1989x}.
The outer radius of accretion disk ($R_{\rm out}$) is frozen at the default value $400R_{\rm g}$ (gravitational radius $R_{\rm g}$ = $GM/c^2$).
Considering the MAXI J0637-430 is a Galactic transient, we fix the redshift ($z$) at zero.
Other parameters like inclination angle ($i$), photon index of the X-ray spectrum ($\Gamma$), ionization of the accretion disk (${\rm log}\xi$), iron abundance ($A_{\rm Fe}$), reflection fraction ($R_{\rm f}$), electron temperature in the corona ($kT_{\rm e}$) and normalization (Norm) vary freely.
The fitting result is shown that the inner radius of the disk extended to the ISCO with $R_{\rm{in}}$ =$ 1.73^{+1.30}_{-0.37} R_{\rm{ISCO}}$.
After examining the location of the inner radius of the disk, we set the $R_{\rm{in}}$ = $R_{\rm{ISCO}}$ and let the black hole spin as a free parameter. 
By using the model \verb'relxillCp' to fit the spectrum of the intermediate state, we obtain that the black hole spin $\textit{a}_{\rm{*}} > 0.72$.
The inclination angle is constrained to be $i$ = $46.1_{-5.3}^{+4.0}$ degrees.
Our fitting result shows that the accretion disk is highly ionized with ${\rm log}\xi$ = $4.30^{+0.17}_{-0.18}$ and a hot corona with $k{T}_{\rm{e}} > 197.1$ keV.
It is worth noting that the fitting result shows a supersolar iron abundance with $\textit{A}_{\rm{Fe}} > 8.89 \textit{A}_{\rm{Fe, \bigodot}}$.
This result has also appeared in other black hole X-ray binaries, and we will discuss the issue of iron abundance in the Section \ref{sec4}.
The spectral fit of model \verb'constant*tbabs*(diskbb+relxillCp)' is shown in Fig. \ref{figre}.
All the best fitting parameters for model \verb'constant*tbabs*(diskbb+relxillCp)' are listed in Table \ref{retab}.

\begin{table*}
\small
  \caption{Fitting results for simple models}
     \label{po}
\begin{threeparttable}      
\begin{tabular}{c@{\hspace{10pt}}c@{\hspace{20pt}}ccccc@{\hspace{10pt}}c@{\hspace{10pt}}c@{\hspace{10pt}}c@{\hspace{10pt}}c@{\hspace{10pt}}c@{\hspace{10pt}}ccc@{\hspace{10pt}}c@{\hspace{10pt}}c}
\toprule
\toprule
\smallskip
\textit{NuSTAR}   & \textit{Swift}/XRT  &  \multicolumn{2}{c}{\tt{diskbb}} & \multicolumn{2}{c}{\tt{powerlaw}}  &  $C_{\rm FPMA}$/$C_{\rm FPMB}$ &Reduced  $\chi_{\nu}^2$  \\
 ObsId&ObsId& $T_{\rm{in}}$& $N_{\rm{diskbb}}$ & $\Gamma$ & $\rm{N_{Powerlaw}}$ $(\times 10^{-1})$&  \\

\hline
80502324002 & -&$0.65_{-0.01}^{+0.01}$ & $1733.6_{-27.6}^{+28.5}$& $2.44_{-0.02}^{+0.02}$ & $1.07_{-0.05}^{+0.06}$  & -/0.99 & 1.80 \\
80502324004 & 00012172008 &$0.62_{-0.01}^{+0.01}$ & $1661.2_{-15.2}^{+15.5}$ & $2.25_{-0.01}^{+0.01}$ & $1.31_{-0.03}^{+0.03}$ & 0.80/0.81 & 1.52  \\
80502324006 &00012172018 & $0.55_{-0.01}^{+0.01}$ & $1584.6_{-24.0}^{+24.6}$ & $2.67_{-0.02}^{+0.03}$ & $0.92_{-0.04}^{+0.05}$ & 0.97/0.94 & $1.46$  \\
80502324008 & 00012172066& $0.17_{-0.01}^{+0.01}$ & $27150.3_{-5095.5}^{+6344.0}$ & $1.86_{-0.01}^{+0.01}$ & $0.26_{-0.01}^{+0.01}$ & 1.15/1.17 & $1.01$   \\
80502324010 &00012172077& - & - & $1.77_{-0.01}^{+0.01}$ & $0.06_{-0.01}^{+0.01}$ & 1.17/1.16 & $1.02$    \\
80502324012 & 00012172085 & - & - & $1.81_{-0.02}^{+0.02}$ & $0.03_{-0.01}^{+0.01}$ & 1.15/1.16 & $0.94$  \\
80502324014& 00012172093& - & - & $1.78_{-0.03}^{+0.03}$ & $0.03_{-0.01}^{+0.01}$ & 1.14/1.16& $0.85$   \\
80502324016& - & - & - & $1.78_{-0.12}^{+0.12}$ & $0.03_{-0.01}^{+0.01}$ & -/1.16 & $0.95$ \\
\bottomrule
\end{tabular}
		\begin{tablenotes}
			\item \textbf{Notes.} The best-fitting parameters obtained by \textit{NuSTAR} observations with model {\tt constant*tbabs*powerlaw} or model {\tt constant*tbabs*(diskbb+powerlaw)}. The parameters with the symbol “$\dagger$” indicate they are fixed at values given. The errors are calculated in 90\% confidence level by \texttt{XSPEC}.
		\end{tablenotes}
\end{threeparttable}    
\end{table*}

\begin{table}
\caption{Best-fitting parameters for relativistic reflection models}
\begin{center}
 \setlength{\tabcolsep}{0.1mm}
			\begin{tabular}{cccccc}
			\toprule
			\toprule
			Components & Parameter & Free $R_{\rm{in}}$ & Free $a_{*}$\\
			 & &$a_{*}=0.998$ &$R_{\rm{in}}$=$R_{\rm{ISCO}}$ & &\\
			\midrule
			\multicolumn{5}{c}{\emph{\textit{NuSTAR}}}\\
			\midrule
			
			\verb'tbabs' & $N_{\rm{H}}$ $(\times$ 10$^{20}$cm$^{-2}$) 
			&$4.39^\dagger$
                &$4.39^\dagger$
			\\
                         [1ex]			
                 \verb'diskbb'
		    &$kT_{\rm{in}}$ (keV)
		    &$0.64_{-0.01}^{+0.01}$
                &$0.64_{-0.01}^{+0.01}$
		        \\
		        [1ex]
		    &$N_{\rm{diskbb}}$
		      &$1908.10_{-37.70}^{+46.28}$
                &$1908.15_{-42.49}^{+44.36}$
			\\
			[1ex]
			
			\verb'relxillCp'

			& $\textit{a}_{\rm{*}}$ 
			&$0.998^\dagger$
                &$0.91_{-0.19}^*$
			\\
                         [1ex]
                & $R_{\rm{in}}$($R_{\rm{ISCO}}$)
                &$1.73_{-0.37}^{+1.30}$
                &$1^\dagger$
                \\
                         [1ex]
			& \textit{i} (deg) 
			&$46.4_{-5.0}^{+3.9}$
                &$46.1_{-5.3}^{+4.0}$
			 \\
                           [1ex]
                & $\Gamma$
			&$2.22_{-0.02}^{+0.03}$
                &$2.22_{-0.03}^{+0.03}$
              \\
                           [1ex]
			& $\textit{A}_{\rm{Fe}}$ 
			&$10.00_{-1.02}^*$ 
                &$10.00_{-1.11}^*$ 
			\\
                         [1ex]
			& $\rm{log}\xi$ 
			&$4.30_{-0.18}^{+0.17}$
                &$4.30_{-0.18}^{+0.17}$
			 \\
                         [1ex]
                & $k{T}_{\rm{e}}$(keV) 
                &$400.0_{-198.2}^*$
			&$400.0_{-202.9}^*$
               \\
                         [1ex]
			& $R_{\rm{ref}}$ 
			&$0.56_{-0.12}^{+0.07}$
                &$0.56_{-0.12}^{+0.16}$
			 \\
                         [1ex]
			& $N_{\rm{relxillCp}}(\times 10^{-4})$ 
			&$5.50_{-0.61}^{+0.42}$
                &$5.49_{-0.85}^{+0.63}$
			 \\
                         [1ex]
			& logN
			&$15.0^\dagger$
                &$15.0^\dagger$
                          \\
                         [1ex]

			\midrule
			
			$C_{\rm FPMB}$ & & 0.99 & 0.99\\
			\midrule
			$\chi^2/\nu$ & & 1235.08/1064 & 1235.82/1064\\
			\midrule
			$\chi_{\nu}^2$ & & 1.16 & 1.16\\
			\bottomrule

			\end{tabular}
		\begin{tablenotes}
			\item \textbf{Notes.} The best-fitting parameters were obtained with reflection model {\tt constant*tbabs*(diskbb+relxillCp)}. The parameters with “$\dagger$” indicate they are fixed at the values given. * indicates that the upper or lower limit of the parameter pegs the maximum or minimum value. The errors are calculated in 90\% confidence level by \texttt{XSPEC}.
		\end{tablenotes}
		\end{center}
	\label{retab}
	\end{table}
\section{Discussion}
\label{sec4}
\subsection{Supersolar iron abundance}
\label{sec4-1}
In this section, we mainly discuss the possible reasons for supersolar iron abundance from reflection analysis of the first observation and the effect of supersolar iron abundance on black hole spin.
In previous studies, some black hole X-ray binaries have shown supersolar iron abundance, such as GX 339-4 ( $\textit{A}_{\rm{Fe}}$ = 5 $\pm$ 1 $\textit{A}_{\rm{Fe, \bigodot}}$ in \citealt{garcia2015x} and $\textit{A}_{\rm{Fe}}$ = 6.6 $\pm$ 0.5 $\textit{A}_{\rm{Fe, \bigodot}}$ in \citealt{parker2016nustar}), V404 Cyg ($\textit{A}_{\rm{Fe}}$ $\sim$ 5 $\textit{A}_{\rm{Fe, \bigodot}}$ in \citealt{walton2017living}), Cyg X-1 ( $\textit{A}_{\rm{Fe}}$ = 4.7 $\pm$ 0.1 $\textit{A}_{\rm{Fe, \bigodot}}$ in \citealt{parker2015nustar} and $\textit{A}_{\rm{Fe}}$ = 4.0 - 4.3 $\textit{A}_{\rm{Fe, \bigodot}}$ in \citealt{walton2016soft}), 4U 1543-47 ( $\textit{A}_{\rm{Fe}}$ = $5.05^{+1.21}_{-0.26}$ $\textit{A}_{\rm{Fe, \bigodot}}$ in \citealt{dong2020spin} and $\textit{A}_{\rm{Fe}}$ 3.6 - 10.0 $\textit{A}_{\rm{Fe, \bigodot}}$ in \citealt{prabhakar2023wideband}), AT2019wey ( $\textit{A}_{\rm{Fe}}$ $\sim$ 5 $\textit{A}_{\rm{Fe, \bigodot}}$ in \citealt{feng2022spectral}), MAXI J1836-194 ( $\textit{A}_{\rm{Fe}}$ $>$ 4.5 $\textit{A}_{\rm{Fe, \bigodot}}$ in \citealt{dong2020detailed}) and MAXI J1348-630 ( $\textit{A}_{\rm{Fe}}$ $\sim$ 7.0 - 10.0 $\textit{A}_{\rm{Fe, \bigodot}}$ in \citealt{jia2022detailed}).
Some of these sources show extreme supersolar iron abundance like MAXI J0637-430.
For the supersolar iron abundance obtained by X-ray reflection, \citet{garcia2018problem} proposes a possible explanation that the model shortfall at very high densities ($n_{\rm{e}}$ $>$ $10^{18}$ $\rm{cm}^{-3}$) due to atomic data shortcomings in this regime. 
In the parameter settings of the old version of the reflection model \verb'relxillCp', the disk density is fixed to $10^{15}$ $\rm{cm}^{-3}$.
However, the prediction of the disk density in the research of the standard $\alpha$-disk model \citep{shakura1973black} and 3D magneto-hydrodynamic (MHD) simulations is 
much larger \citep{noble2010dependence,schnittman2013x}.
The higher density of the accretion disk will contribute to the spectra in two aspects, 1) free-free heating produces a flux excess at soft energies; 2) effect on the atomic parameters control line emission and photoelectric absorption. 
The underestimation of the disk density may lead to the issue of supersolar iron abundance.
The latest version of the reflection model \verb'relxillCp' already allows free fitting of the disk density, ranging from $10^{15}$ $\rm{cm}^{-3}$ to $10^{20}$ $\rm{cm}^{-3}$.
In the previous work of \citet{jia2022detailed}, the high-density model is successfully used to explain the high iron abundance and the fitting results show that it has a negligible effect on the spin measurement.
Therefore, on the basis of the model \verb'constant*tbabs*(diskbb+relxillCp)' in Section \ref{sec3}, the disk density is freely fitted.
We set the inclination angle to range from 40.8 - 50.1 degrees which is obtained by fitting spectrum in Section \ref{sec3}.
From the fitting results by using high-density reflection model \verb'constant*tbabs*(diskbb+relxillCp)', we can see that the disk density will be pegged at the maximum value of $10^{20}$ $\rm{cm}^{-3}$ after free fitting, and the iron abundance appears to be much larger than that of the solar abundance.
The phenomenon may be caused by the maximum disk density ($10^{20}$ $\rm{cm}^{-3}$) of the model \verb'relxillCp' is still not enough to fit the actual value of the iron abundance.

\citet{tomsick2018alternative} obtained the results of extreme supersolar iron abundance when using the reflection model of constant density ($10^{15}$ $\rm{cm}^{-3}$) to study Cyg X-1.
They use a new version of reflection model \verb'reflionx_hd'\footnote{https://ftp.ast.cam.ac.uk/pub/mlparker/reflionx/}  to fit the spectrum, which obtained a result of high disk density $n_{\rm{e}}$ = $3.98 \times 10^{20}$ $\rm{cm}^{-3}$\citep{tomsick2018alternative}.
This decreases the need for extremely supersolar abundances.
We combine the relativistic convolution model \verb'relconv' and the high disk density model \verb'reflionx_hd' as the reflection component.  Moreover, we plus a Comptonization continuum model \verb'nthcomp' in the whole model configuration.
In the parameters setting, we tie the photon index $\Gamma$ of \verb'reflionx_hd' to the photon index $\Gamma$ of \verb'nthcomp'.
The inclination angle of the accretion disk is in the range of 40.8 - 50.1 degrees ( obtained from fitting results of Section \ref{sec3} ).
Because the temperature of the corona is so high that it cannot be limited, we fix it at the maximum.
Additionally, the iron abundance of this model is set at the solar abundance.
Under the assumption that the iron abundance is set at the solar abundance, a higher disk density ( $n_{\rm{e}} > 2.34 \times 10^{21}$ $\rm{cm}^{-3}$ ) is obtained by fitting.
This result suggests that MAXI J0637-430 is a stellar-mass black hole with an high-density accretion disk, which compared to the typical black hole X-ray binary systems .
There is no significant difference in the spin value obtained by using the high-density model ( both \verb'relxillCp' and \verb'reflionx_hd' ).
For model \verb'relxillCp', the black hole spin obtained by fitting is 
$\textit{a}_{\rm{*}} > 0.83$.
For model \verb'reflionx_hd', the black hole spin obtained by fitting is 
$\textit{a}_{\rm{*}} > 0.79$.
In section \ref{sec3}, we obtain the spin of MAXI J0637-430 with $\textit{a}_{\rm{*}} > 0.72$, which shows that the changes of iron abundance and accretion disk density have a negligible effect on the spin measurement.

In addition, there is another possibility to explain the extremely supersolar iron abundance.
According to the research of \citet{kinch2021spin} and \citet{mondal2021emission}, the Fe K$\alpha$ line strength increases with Fe abundance for sub-linearly.
Therefore, an iron line with higher equivalent width may lead to an abnormal increase in iron abundance.
When we use the model \verb'constant*tbabs*(diskbb+gaussian+powerlaw)' for spectral fitting, we calculate the equivalent width of the model \verb'gaussian' used to fit the iron line component.
The equivalent width is about 381 eV, which represents a relatively strong iron line component.
Besides, the line width of the \verb'gaussian' component is up to 1.4 keV.
Therefore, the iron line profile is highly broadened by the strong gravity.
Such a strong iron line may be a potential reason for the extremely high iron abundance.
The discovery is expected to be verified in more researches in the future.

\subsection{Soft excess}
\label{soex}
In the research of \citet{lazar2021spectral}, the soft excess may be an emission from a combination of thermal Comptonization component and reflection component of disk blackbody returning radiation.
We also explored this possible scenario by using the reflection model \verb'relxillNS'\footnote{http://www.sternwarte.uni-erlangen.de/~dauser/research/relxill/}, in which the incident spectrum was changed to a blackbody radiation.
This model is usually used to study the reflection components in the radiation of neutron stars.
We set the blackbody temperature $kT_{\rm{bb}}$ in the reflection model \verb'relxillNS' equal to the temperature of the disk component \verb'diskbb', which represents the case that the light bent back to the accretion disk by strong gravity.
In addition, we set the black hole spin and accretion disk inclination angle to the results in Section \ref{sec3}.
The fitting results of the soft state spectra by using \verb'relxillNS' model are shown in Table \ref{diffre}.
The spectra can be well fitted by using the black hole spin and inclination results that we obtained in Section \ref{sec3}.
The additional thermal residuals  in the soft state can be explained by the returning blackbody radiation, which is consistent with the conclusion in \citet{lazar2021spectral}.
In Section \ref{sec4-1}, we also discussed that the soft excess in the intermediate state spectrum may be caused by the large density of the accretion disk. 
Therefore, we attempt to use the high-density version of the typical reflection model \verb'relxillCp' to fit the soft state spectral, which the incident radiation is a Comptonization component from the hot corona.
The fitting results of the soft state spectra by using \verb'relxillCp' model are shown in Table \ref{diffre}.
We find that \verb'relxillCp' model can also obtain an acceptable fitting statistical results.
Comparing the fitting results of the two reflection models, we find that the blackbody temperature of the accretion disk is lower when using the \verb'relxillNS' model, while the higher disk density is obtained when using the \verb'relxillCp' model, which may be due to different description of the mechanism of soft excess.
From the fitting statistical results of the two reflection models, the \verb'relxillNS' model can provide a better fitting.
To distinguish which physical scenario is more physical may require more observations and studies in the future.
At present, both the high density accretion disk and the return radiation from the accretion disk may be the potential reasons.

\begin{table}
\caption{Best-fitting parameters for different incident radiation of the reflection model}
\begin{center}
 \setlength{\tabcolsep}{0.5mm}
			\begin{tabular}{ccccccc}
			\toprule
			\toprule
			Components & Parameter &  \multicolumn{2}{c}{MJD 58801}&\multicolumn{2}{c}{MJD 58812}\\
			 & & & &\\
			\midrule
			\multicolumn{5}{c}{\emph{\textit{NuSTAR}-\textit{Swift}/\rm{XRT}}}\\
			\midrule
			
			\verb'tbabs' & $N_{\rm{H}}$ $(\times$ 10$^{20}$cm$^{-2}$) 
			&$4.39^\dagger$
                &$4.39^\dagger$
                
                 &$4.39^\dagger$
                  &$4.39^\dagger$
			\\
                         [1ex]			
                 \verb'diskbb'
		    &$kT_{\rm{in}}$ (keV)
		    &$0.56_{-0.01}^{+0.01}$
                &$0.62_{-0.01}^{+0.01}$
                 
                 &$0.53_{-0.01}^{+0.01}$
                  &$0.55_{-0.01}^{+0.01}$
		        \\
		        [1ex]
		    &$N_{\rm{diskbb}}$
		      &$1489.87_{-37.70}^{+46.28}$
                &$1755.03_{-16.92}^{+19.94}$
                 
                &$1477.52_{-51.27}^{+40.74}$
                &$1524.27_{-42.49}^{+44.36}$
			\\
			[1ex]
			 \verb'nthcomp'                
                & $\Gamma$
		&$2.07_{-0.03}^{+0.03}$
                &$2.15_{-0.03}^{+0.03}$
                
                &$2.32_{-0.02}^{+0.03}$
                &$2.36_{-0.03}^{+0.03}$
             \\
                         [1ex]
                 & $k{T}_{\rm{e}}$(keV) 
                &>330.6
			&>437.4
			&>219.8
			&>188.1
               \\
                         [1ex]
                  & $N_{\rm{nthcomp}}(\times 10^{-2})$
			&$1.65_{-0.61}^{+1.98}$
                &$1.24_{-0.01}^{+0.01}$
                &$0.72_{-0.03}^{0.05}$
                &$0.32_{-0.01}^{+0.08}$
                   \\
                         [1ex]
			\verb'relxillNS'/\verb'relxillCp'

			& $\textit{a}_{\rm{*}}$ 
			&$0.91^\dagger$
                &$0.91^\dagger$
                &$0.91^\dagger$
                &$0.91^\dagger$
			\\
                         [1ex]
                & $kT_{\rm{bb}}^*$(keV)
                &$0.56_{-0.01}^{+0.01}$
                &$-$
                &$0.53_{-0.01}^{+0.01}$
                &$-$
                \\
                         [1ex]
			& \textit{i} (deg) 
			&$46.1^\dagger$
                &$46.1^\dagger$
                &$46.1^\dagger$
                &$46.1^\dagger$
			 \\
                           [1ex]
			& $\textit{A}_{\rm{Fe}}$ 
			&$0.50^{+0.01}$ 
                &$1.67_{-0.60}^{+0.60}$ 
                &$0.50^{+0.08}$
                &$0.72_{-0.15}^{+0.32}$ 
			\\
                         [1ex]
			& $\rm{log}\xi$ 
			&$2.92_{-0.02}^{+0.02}$
                &$2.89_{-0.06}^{+0.10}$
                &$2.27_{-0.09}^{+0.08}$
                &$2.66_{-0.08}^{+0.07}$
			 \\
                         [1ex]
			& logN
			&$19.0_{-0.6}$
                &$19.0_{-0.1}^{+0.1}$
               & $18.0_{-0.4}^{+0.1}$
               &$19.0_{-0.1}^{+0.1}$
                          \\
                         [1ex]
& $N_{\rm{relxillNS}}(\times 10^{-3})$ 
			&$3.48_{-0.24}^{+0.43}$
                &$-$
                &$1.97_{-0.15}^{+0.16}$
                 &$-$
                           \\
                           [1ex]
                           & $N_{\rm{relxillCp}}(\times 10^{-4})$ 
			&$-$
                &$4.52_{-0.48}^{+0.56}$
                 &$-$
                 &$4.37_{-0.56}^{+0.50}$
                           \\
                           [1ex]
			\midrule
			
			$C_{\rm FPMA}/C_{\rm FPMB}$ & & 0.84/0.85 & 0.80/0.81& 0.99/0.96& 0.96/0.94\\
			\midrule
			$\chi^2/\nu$ & & 2536.13/2206 & 2686.52/2206 &1725.76/1511  &1782.54/1511\\
			\midrule
			$\chi_{\nu}^2$ & & 1.15 & 1.21 & 1.14& 1.18\\
			\bottomrule

			\end{tabular}
		\begin{tablenotes}
			\item \textbf{Notes.} The best-fitting parameters were obtained with different incident radiation of the reflection model. The parameters with “$\dagger$” indicate they are fixed at the values given. * indicates that the blackbody temperature of the {\tt{relxillNS}} is set equal to the temperature of the  {\tt{diskbb}}. The errors are calculated in 90\% confidence level by \texttt{XSPEC}.
		\end{tablenotes}
		\end{center}
	\label{diffre}
	\end{table}

\subsection{Spin parameter}
\label{sec4-2}
Using the most state-of-art reflection model to fit the spectral, the spin parameter of MAXI J0637-430 is measured with $\textit{a}_{\rm{*}} > 0.72$.
In a recent work of MAXI J0637-430, \citet{soria20222} proposed to calculate black hole spin $(\textit{a}_{\rm{*}} < 0.25$) using mass and distance.
This inconsistency may be caused by  the difference between the inclination angle and the inner radius of the disk.
Under this method, the black hole mass, source distance, and inclination angle of the accretion disk have a great effect on the spin measurement.
Usually, we use the continuum-fitting method to measure the black hole spin when the dynamical parameters are known.
Using the X-ray reflection spectroscopy, we only need to obtain the black hole spin by fitting the reflection components in the spectral regardless of the precise dynamical parameters. 
In the work of \citet{kinch2021spin} mentioned in Section \ref{sec4-1}, the Fe K$\alpha$ profile is more sensitive to the accretion rate than to the black hole spin.
According to the \citet{tetarenko2021using} and \citet{ma2022peculiar}, MAXI J0637-430 has a relatively low accretion rate. 
This may suggest that the broadening of the Fe K$\alpha$ emission line is caused by the strong gravitational redshift effect from relatively high black hole spin rather than the mass accretion rate.

\begin{table}
\caption{Best-fitting parameters for high-density relativistic reflection models}
\begin{center}
 \setlength{\tabcolsep}{0.1mm}
			\begin{tabular}{cccccc}
			\toprule
			\toprule
			Components & Parameter & M1 & M2\\
			 & & & & &\\
			\midrule
			\multicolumn{5}{c}{\emph{\textit{NuSTAR}}}\\
			\midrule
			
			\verb'tbabs' & $N_{\rm{H}}$ $(\times$ 10$^{20}$cm$^{-2}$) 
			&$4.39^\dagger$
                &$4.39^\dagger$
			\\
                         [1ex]			
                 \verb'diskbb'
		    &$kT_{\rm{in}}$ (keV)
		    &$0.65_{-0.01}^{+0.01}$
                &$0.64_{-0.01}^{+0.01}$
		        \\
		        [1ex]
		    &$N_{\rm{diskbb}}$
		      &$1871.83_{-34.50}^{+33.12}$
                &$2005.88_{-42.15}^{+46.43}$
			\\
			[1ex]
			
			\verb'relxillCp'

			& $\textit{a}_{\rm{*}}$ 
			&$0.92_{-0.09}^*$
                &$-$
                \\
                         [1ex]
			& \textit{i} (deg) 
			&$40.8_*^{+4.0}$
                &$-$
			 \\
                           [1ex]
                & $\Gamma$
			&$2.07_{-0.04}^{+0.07}$
                &$-$
              \\
                           [1ex]
			& $\textit{A}_{\rm{Fe}}$ 
			&$10.00_{-1.33}^*$ 
                &$-$ 
			\\
                         [1ex]
			& $\rm{log}\xi$ 
			&$3.77_{-0.96}^{+0.13}$
                &$-$
			 \\
                         [1ex]
                & $k{T}_{\rm{e}}$(keV) 
                &$400.0_{-216.0}^*$
			&$-$
               \\
                         [1ex]
			& $R_{\rm{ref}}$ 
			&$0.62_{-0.17}^{+0.37}$
                &$-$
			 \\
                         [1ex]
			& $N_{\rm{relxillCp}}(\times 10^{-4})$ 
			&$3.32_{-0.81}^{+0.91}$
                &$-$
			 \\
                         [1ex]
			& logN
			&$19.5_{-1.2}^*$
                &$-$
                          \\
                         [1ex]
                \verb'nthcomp'                
                & $\Gamma$
			&$-$
                &$1.94_{-0.03}^{+0.03}$
             \\
                         [1ex]
                  & $N_{\rm{nthcomp}}(\times 10^{-2})$
			&$-$
                &$1.91_{-0.61}^{+1.98}$
                   \\
                         [1ex]
                \verb'relconv'
                & $\textit{a}_{\rm{*}}$
                &$-$
                &$0.93_{-0.14}^*$
			\\
                         [1ex]
			& \textit{i} (deg) 
			&$-$
                &$40.8^{+2.5}_*$
			 \\
                           [1ex]
                \verb'reflionx_hd'
			& $\textit{A}_{\rm{Fe}}$ 
			&$-$ 
                &$1^\dagger$ 
			\\
                         [1ex]
			& $\rm{log}\xi$ 
			&$-$
                &$3.16_{-0.06}^{+0.25}$
               \\
                         [1ex]
			& $N_{\rm{reflionx\_hd}} (\times 10^{-2})$ 
			&$-$
                &$8.99_{-0.94}^{+0.84}$
			 \\
                         [1ex]
			& logN
			&$-$
                &$22.0_{-0.6}^*$
                \\
                [1ex]
			\midrule
			
			$C_{\rm FPMB}$ & & 0.99 & 0.99\\
			\midrule
			$\chi^2/\nu$ & & 1229.01/1063 & 1181.94/1065\\
			\midrule
			$\chi_{\nu}^2$ & & 1.15 & 1.11\\
			\bottomrule

			\end{tabular}
		\begin{tablenotes}
			\item \textbf{Notes.} The best-fit parameters were obtained by fitting the \textit{NuSTAR} spectra for high disk density reflection model M1: {\tt constant*tbabs*(diskbb+relxillCp)} and M2: {\tt constant*tbabs*(diskbb+nthcomp+relconv*reflionx\_hd)}. The parameters with “$\dagger$” indicate they are fixed at the values given.  * indicates that the upper or lower limit of the parameter pegs the maximum or minimum value. The errors are calculated in 90\% confidence level by \texttt{XSPEC}.
		\end{tablenotes}
		\end{center}
	\label{hiretab}
	\end{table}

\section{conclusion}
\label{sec5}
 In this work, we mainly analyze the \textit{NuSTAR} and \textit{Swift}/XRT archived data of MAXI J0637-430 and use the spectra with strong reflection components for spin measurement.
 The reflection model \verb'relxillCp' can be used to fit the spectra well, and the parameters characterized by physical properties can be obtained, including the relatively high black hole spin $\textit{a}_{\rm{*}} > 0.72$ and the inclination angle of the accretion disk $i$ = $46.1_{-5.3}^{+4.0}$ degrees (at 90\% confidence level). 
Remarkably, the fitting results show the extremely high iron abundance in the intermediate state.
Using the reflection model with higher disk density, we get a high-density accretion disk ( $n_{\rm{e}} > 2.34 \times 10^{21}$ $\rm{cm}^{-3}$ ) under the assumption of a solar abundance, which is consistent with the results predicted by previous magnetohydrodynamics simulations.
In addition, the relatively strong Fe k$\alpha$ line, that is, the Fe K$\alpha$ line with higher equivalent width, will show a sub-linear relationship with iron abundance.
The equivalent width of the Fe K$\alpha$ line of MAXI J0637-430 is about 381 eV, which may be the potential reason for the extreme supersolar iron abundance.
Moreover, we discussed the results from \citet{lazar2021spectral}.
We prove that the black hole spin and inclination angle obtained by fitting can be used to well describe the soft state spectra.
The additional residuals shown in the soft state spectral can be well fitted by either the model \verb'relxillNS' or the high density version of the model \verb'relxillCp '.
This suggests that there are two possible scenarios to explain the soft excess in the thermal state, one is from \citet{lazar2021spectral}, which the soft excess is from a combination of thermal Comptonization component and reflection component of disk blackbody returning radiation, a light bending effect caused by the strong gravity.
The other is from the extra free-free heating caused by the high density of the accretion disk and produces a flux excess at soft energies.
At present, the study cannot determine which physical scenario is more realistic, and we hope that more data and more advanced models can confirm the difference between the two possibilities in the future.
The spin parameter is also discussed, and its reliability is verified from the aspects of the model fitting and theoretical application.
The iron abundance and disk density have a negligible effect on the spin measurement results.

\section*{Acknowledgements}
We thank the anonymous referee for useful comments. This work has made use of data obtained from the \textit{NuSTAR} satellite, a Small Explorer mission led by the California Institute of Technormlogy (Caltech) and managed by NASA's Jet Propulsion Laboratory in Pasadena. 
We thank the \textit{NuSTAR} Operations, Software, and Calibration teams for support with the execution and analysis of these observations.
This research has made use of the \textit{NuSTAR} Data Analysis Software NuSTARDAS, jointly developed by the ASI Science Data Center (ASDC, Italy) and the California Institute of Technormlogy (USA).
This work made use of data supplied by UK Swift Science Data Centre.
L.J.G. is supported by the National Natural Science Foundation of China (Grant No. U12273058).

\section*{Data Availability}
The data underlying this article are observed by \textit{NuSTAR} which is accessed from 
\noindent \url{https://heasarc.gsfc.nasa.gov/xamin} 


\bibliographystyle{raa}
\bibliography{bibtex}

\begin{thebibliography}{70}
\providecommand\natexlab[1]{#1}
\providecommand\JournalTitle[1]{#1}

\bibitem[Baby {et~al.}(2021)]{baby2021revealing}
Baby, B.~E., Bhuvana, G., Radhika, D., {et~al.} 2021, Monthly Notices of the
  Royal Astronomical Society, 508, 2447

\bibitem[Bardeen {et~al.}(1972)]{bardeen1972rotating}
Bardeen, J.~M., Press, W.~H., \& Teukolsky, S.~A. 1972, Astrophysical Journal,
  Vol. 178, pp. 347-370 (1972), 178, 347

\bibitem[Burrows {et~al.}(2005)]{burrows2005swift}
Burrows, D.~N., Hill, J., Nousek, J., {et~al.} 2005, Space science reviews,
  120, 165

\bibitem[Dauser {et~al.}(2014)]{dauser2014role}
Dauser, T., Garc{\'\i}a, J., Parker, M., Fabian, A., \& Wilms, J. 2014, Monthly
  Notices of the Royal Astronomical Society: Letters, 444, L100

\bibitem[Dauser {et~al.}(2013)]{dauser2013irradiation}
Dauser, T., Garcia, J., Wilms, J., {et~al.} 2013, Monthly Notices of the Royal
  Astronomical Society, 430, 1694

\bibitem[Dauser {et~al.}(2010)]{dauser2010broad}
Dauser, T., Wilms, J., Reynolds, C., \& Brenneman, L. 2010, Monthly Notices of
  the Royal Astronomical Society, 409, 1534

\bibitem[Dong {et~al.}(2020{\natexlab{a}})]{dong2020detailed}
Dong, Y., Garc{\'\i}a, J.~A., Liu, Z., {et~al.} 2020{\natexlab{a}}, Monthly
  Notices of the Royal Astronomical Society, 493, 2178

\bibitem[Dong {et~al.}(2020{\natexlab{b}})]{dong2020spin}
Dong, Y., Garc{\'\i}a, J.~A., Steiner, J.~F., \& Gou, L. 2020{\natexlab{b}},
  Monthly Notices of the Royal Astronomical Society, 493, 4409

\bibitem[Dong {et~al.}(2022)]{dong2022analysis}
Dong, Y., Liu, Z., Tuo, Y., {et~al.} 2022, Monthly Notices of the Royal
  Astronomical Society, 514, 1422

\bibitem[Evans {et~al.}(2009)]{evans2009methods}
Evans, P., Beardmore, A., Page, K., {et~al.} 2009, Monthly Notices of the Royal
  Astronomical Society, 397, 1177

\bibitem[Fabian {et~al.}(1989)]{fabian1989x}
Fabian, A., Rees, M., Stella, L., \& White, N.~E. 1989, Monthly Notices of the
  Royal Astronomical Society, 238, 729

\bibitem[Feng {et~al.}(2022{\natexlab{a}})]{feng2022using}
Feng, Y., Steiner, J.~F., Ramirez, S.~U., \& Gou, L. 2022{\natexlab{a}}, arXiv
  preprint arXiv:2212.04653

\bibitem[Feng {et~al.}(2022{\natexlab{b}})]{feng2022spin}
Feng, Y., Zhao, X., Li, Y., {et~al.} 2022{\natexlab{b}}, Monthly Notices of the
  Royal Astronomical Society, 516, 2074

\bibitem[Feng {et~al.}(2022{\natexlab{c}})]{feng2022estimating}
Feng, Y., Zhao, X., Gou, L., {et~al.} 2022{\natexlab{c}}, The Astrophysical
  Journal, 925, 142

\bibitem[Feng {et~al.}(2022{\natexlab{d}})]{feng2022spectral}
Feng, Y., Zhao, X., Gou, L., {et~al.} 2022{\natexlab{d}}, Science China
  Physics, Mechanics \& Astronomy, 65, 1

\bibitem[Garc{\'\i}a {et~al.}(2015)]{garcia2015x}
Garc{\'\i}a, J.~A., Steiner, J.~F., McClintock, J.~E., {et~al.} 2015, The
  Astrophysical Journal, 813, 84

\bibitem[Garc{\'\i}a {et~al.}(2018{\natexlab{a}})]{garcia2018reflection}
Garc{\'\i}a, J.~A., Steiner, J.~F., Grinberg, V., {et~al.} 2018{\natexlab{a}},
  The Astrophysical Journal, 864, 25

\bibitem[Garc{\'\i}a {et~al.}(2013)]{garcia2013x}
Garc{\'\i}a, J., Dauser, T., Reynolds, C., {et~al.} 2013, The Astrophysical
  Journal, 768, 146

\bibitem[Garc{\'\i}a {et~al.}(2018{\natexlab{b}})]{garcia2018problem}
Garc{\'\i}a, J., Kallman, T., Bautista, M., {et~al.} 2018{\natexlab{b}}, arXiv
  preprint arXiv:1805.00581

\bibitem[Garc{\'\i}a {et~al.}(2011)]{garcia2011x}
Garc{\'\i}a, J., Kallman, T., \& Mushotzky, R. 2011, The Astrophysical Journal,
  731, 131

\bibitem[Garcia \& Kallman(2010)]{garcia2010x}
Garcia, J., \& Kallman, T.~R. 2010, The Astrophysical Journal, 718, 695

\bibitem[Garc{\'\i}a {et~al.}(2014)]{garcia2014improved}
Garc{\'\i}a, J., Dauser, T., Lohfink, A., {et~al.} 2014, The Astrophysical
  Journal, 782, 76

\bibitem[Gendreau {et~al.}(2012)]{gendreau2012neutron}
Gendreau, K.~C., Arzoumanian, Z., \& Okajima, T. 2012, in Space Telescopes and
  Instrumentation 2012: Ultraviolet to Gamma Ray, Vol. 8443, SPIE, 322

\bibitem[Gou {et~al.}(2010)]{gou2010spin}
Gou, L., McClintock, J.~E., Steiner, J.~F., {et~al.} 2010, The Astrophysical
  Journal Letters, 718, L122

\bibitem[Gou {et~al.}(2009)]{gou2009determination}
Gou, L., McClintock, J.~E., Liu, J., {et~al.} 2009, The Astrophysical Journal,
  701, 1076

\bibitem[Gou {et~al.}(2014)]{gou2014confirmation}
Gou, L., McClintock, J.~E., Remillard, R.~A., {et~al.} 2014, The Astrophysical
  Journal, 790, 29

\bibitem[Harrison {et~al.}(2013)]{harrison2013nuclear}
Harrison, F.~A., Craig, W.~W., Christensen, F.~E., {et~al.} 2013, The
  Astrophysical Journal, 770, 103

\bibitem[Jana {et~al.}(2021)]{jana2021nicer}
Jana, A., Jaisawal, G.~K., Naik, S., {et~al.} 2021, Monthly Notices of the
  Royal Astronomical Society, 504, 4793

\bibitem[Jia {et~al.}(2022)]{jia2022detailed}
Jia, N., Zhao, X., Gou, L., {et~al.} 2022, Monthly Notices of the Royal
  Astronomical Society, 511, 3125

\bibitem[Kaastra \& Bleeker(2016)]{kaastra2016optimal}
Kaastra, J., \& Bleeker, J. 2016, Astronomy \& Astrophysics, 587, A151

\bibitem[Kennea {et~al.}(2019)]{kennea2019maxi}
Kennea, J., Bahramian, A., Evans, P., {et~al.} 2019, The Astronomer's Telegram,
  13257, 1

\bibitem[Kinch {et~al.}(2021)]{kinch2021spin}
Kinch, B.~E., Schnittman, J.~D., Noble, S.~C., Kallman, T.~R., \& Krolik, J.~H.
  2021, The Astrophysical Journal, 922, 270

\bibitem[Kolehmainen \& Done(2010)]{kolehmainen2010limits}
Kolehmainen, M., \& Done, C. 2010, Monthly Notices of the Royal Astronomical
  Society, 406, 2206

\bibitem[Lazar {et~al.}(2021)]{lazar2021spectral}
Lazar, H., Tomsick, J.~A., Pike, S.~N., {et~al.} 2021, The Astrophysical
  Journal, 921, 155

\bibitem[Ma {et~al.}(2022)]{ma2022peculiar}
Ma, R., Soria, R., Tao, L., {et~al.} 2022, Monthly Notices of the Royal
  Astronomical Society, 514, 5238

\bibitem[Matsuoka {et~al.}(2009)]{matsuoka2009maxi}
Matsuoka, M., Kawasaki, K., Ueno, S., {et~al.} 2009, Publications of the
  Astronomical Society of Japan, 61, 999

\bibitem[Miller {et~al.}(2013)]{miller2013nustar}
Miller, J., Parker, M., Fuerst, F., {et~al.} 2013, The Astrophysical Journal
  Letters, 775, L45

\bibitem[Mondal {et~al.}(2021)]{mondal2021emission}
Mondal, S., Adhikari, T.~P., \& Singh, C.~B. 2021, Monthly Notices of the Royal
  Astronomical Society, 505, 1071

\bibitem[Noble {et~al.}(2010)]{noble2010dependence}
Noble, S.~C., Krolik, J.~H., \& Hawley, J.~F. 2010, The Astrophysical Journal,
  711, 959

\bibitem[Orosz {et~al.}(2002)]{orosz2002dynamical}
Orosz, J.~A., Groot, P.~J., van~der Klis, M., {et~al.} 2002, The Astrophysical
  Journal, 568, 845

\bibitem[Orosz {et~al.}(2007)]{orosz200715}
Orosz, J.~A., McClintock, J.~E., Narayan, R., {et~al.} 2007, Nature, 449, 872

\bibitem[Parker {et~al.}(2015)]{parker2015nustar}
Parker, M., Tomsick, J., Miller, J., {et~al.} 2015, The Astrophysical Journal,
  808, 9

\bibitem[Parker {et~al.}(2016)]{parker2016nustar}
Parker, M., Tomsick, J., Kennea, J., {et~al.} 2016, The Astrophysical Journal
  Letters, 821, L6

\bibitem[Prabhakar {et~al.}(2023)]{prabhakar2023wideband}
Prabhakar, G., Mandal, S., GR, B., \& Nandi, A. 2023, Monthly Notices of the
  Royal Astronomical Society

\bibitem[Reid {et~al.}(2014)]{reid2014parallax}
Reid, M., McClintock, J., Steiner, J., {et~al.} 2014, The Astrophysical
  Journal, 796, 2

\bibitem[Reis {et~al.}(2009)]{reis2009determining}
Reis, R., Fabian, A., Ross, R., \& Miller, J. 2009, Monthly Notices of the
  Royal Astronomical Society, 395, 1257

\bibitem[Remillard \& McClintock(2006)]{remillard2006x}
Remillard, R.~A., \& McClintock, J.~E. 2006, Annu. Rev. Astron. Astrophys., 44,
  49

\bibitem[Reynolds(2021)]{reynolds2021observational}
Reynolds, C.~S. 2021, Annual Review of Astronomy and Astrophysics, 59, 117

\bibitem[Schnittman {et~al.}(2013)]{schnittman2013x}
Schnittman, J.~D., Krolik, J.~H., \& Noble, S.~C. 2013, The Astrophysical
  Journal, 769, 156

\bibitem[Shafee {et~al.}(2006)]{shafee2006estimating}
Shafee, R., McClintock, J., Narayan, R., {et~al.} 2006, AAS/High Energy
  Astrophysics Division\# 9, 9, 1

\bibitem[Shakura \& Sunyaev(1973)]{shakura1973black}
Shakura, N.~I., \& Sunyaev, R.~A. 1973, Astronomy and Astrophysics, 24, 337

\bibitem[Sharma {et~al.}(2019)]{sharma2019study}
Sharma, R., Jain, C., \& Dutta, A. 2019, Monthly Notices of the Royal
  Astronomical Society, 482, 1634

\bibitem[Singh {et~al.}(2014)]{singh2014astrosat}
Singh, K.~P., Tandon, S., Agrawal, P., {et~al.} 2014, in Space Telescopes and
  Instrumentation 2014: Ultraviolet to Gamma Ray, Vol. 9144, SPIE, 517

\bibitem[Soria {et~al.}(2022)]{soria20222}
Soria, R., Ma, R., Tao, L., \& Zhang, S.-N. 2022, Monthly Notices of the Royal
  Astronomical Society, 515, 3105

\bibitem[Steiner {et~al.}(2011)]{steiner2011spin}
Steiner, J.~F., Reis, R.~C., McClintock, J.~E., {et~al.} 2011, Monthly Notices
  of the Royal Astronomical Society, 416, 941

\bibitem[Steiner {et~al.}(2012)]{steiner2012broad}
Steiner, J.~F., Reis, R.~C., Fabian, A.~C., {et~al.} 2012, Monthly Notices of
  the Royal Astronomical Society, 427, 2552

\bibitem[Tetarenko {et~al.}(2021)]{tetarenko2021using}
Tetarenko, B., Shaw, A., Manrow, E., {et~al.} 2021, Monthly Notices of the
  Royal Astronomical Society, 501, 3406

\bibitem[Tomsick {et~al.}(2013)]{tomsick2013reflection}
Tomsick, J.~A., Nowak, M.~A., Parker, M., {et~al.} 2013, The Astrophysical
  Journal, 780, 78

\bibitem[Tomsick {et~al.}(2018)]{tomsick2018alternative}
Tomsick, J.~A., Parker, M.~L., Garc{\'\i}a, J.~A., {et~al.} 2018, The
  Astrophysical Journal, 855, 3

\bibitem[Verner {et~al.}(1996)]{verner1996atomic}
Verner, D., Ferland, G.~J., Korista, K., \& Yakovlev, D. 1996, arXiv preprint
  astro-ph/9601009

\bibitem[Walton {et~al.}(2016)]{walton2016soft}
Walton, D., Tomsick, J., Madsen, K., {et~al.} 2016, The Astrophysical Journal,
  826, 87

\bibitem[Walton {et~al.}(2017)]{walton2017living}
Walton, D., Mooley, K., King, A., {et~al.} 2017, The Astrophysical Journal,
  839, 110

\bibitem[Wang-Ji {et~al.}(2018)]{wang2018evolution}
Wang-Ji, J., Garc{\'\i}a, J.~A., Steiner, J.~F., {et~al.} 2018, The
  Astrophysical Journal, 855, 61

\bibitem[Wilms {et~al.}(2000)]{wilms2000absorption}
Wilms, J., Allen, A., \& McCray, R. 2000, The Astrophysical Journal, 542, 914

\bibitem[Xu {et~al.}(2018)]{xu2018reflection}
Xu, Y., Harrison, F.~A., Garc{\'\i}a, J.~A., {et~al.} 2018, The Astrophysical
  Journal Letters, 852, L34

\bibitem[Zhang {et~al.}(1997)]{zhang1997black}
Zhang, S.~N., Cui, W., \& Chen, W. 1997, The Astrophysical Journal, 482, L155

\bibitem[Zhang {et~al.}(2020)]{apjlac7711bib58}
Zhang, S.-N., Li, T., Lu, F., {et~al.} 2020, SCPMA, 63

\bibitem[Zhao {et~al.}(2020)]{zhao2020confirming}
Zhao, X.-S., Dong, Y.-T., Gou, L.-J., {et~al.} 2020, Journal of High Energy
  Astrophysics, 27, 53

\bibitem[Zhao {et~al.}(2021{\natexlab{a}})]{zhao2021estimating}
Zhao, X., Gou, L., Dong, Y., {et~al.} 2021{\natexlab{a}}, The Astrophysical
  Journal, 916, 108

\bibitem[Zhao {et~al.}(2021{\natexlab{b}})]{zhao2021re}
Zhao, X., Gou, L., Dong, Y., {et~al.} 2021{\natexlab{b}}, The Astrophysical
  Journal, 908, 117

\end{thebibliography}

\end{document}